\documentclass[12pt]{iopart}

\eqnobysec
\usepackage{iopams}

\usepackage{graphicx}

\usepackage{amssymb}

\newtheorem{teo}{Theorem}
\newtheorem{prop}{Proposition}
\newtheorem{cor}[teo]{Corollary}
\newtheorem{lema}[teo]{Lemma}

\newcommand{\CC}{\mathbb{C}}
\newcommand{\RR}{\mathbb{R}}
\newcommand{\ZZ}{\mathbb{Z}}
\newcommand{\NN}{\mathbb{N}}
\newcommand{\Ac}{\mathcal{A}}

\newcommand{\Hc}{\mathcal{H}}
\newcommand{\Tc}{\mathcal{T}}
\newcommand{\Lc}{\mathcal{L}}

\newcommand{\Uc}{\mathcal{U }}
\newcommand{\rank}{\mathrm{rank}}

\newcommand{\bra}[1]{|{#1}\rangle}

\begin{document}

\title{Towards a quantum sampling theory: the case of finite groups}

\author{Antonio~G. Garc\'{\i}a$^{1}$, Miguel~A. Hern\'andez-Medina$^{2}$, A. Ibort$^{1,3}$}

\address{$^1$Department of Mathematics, Universidad Carlos III de Madrid, Av. Universidad 30
28911 Legan\'es, Madrid, Spain.}
\address{$^2$Departamento de Matem\'atica Aplicada a las Tecnolog\'{\i}as de la Informaci\'on y las Comunicaciones, E.T.S.I.T., U.P.M.,
 Avda. Complutense 30, 28040 Madrid, Spain.}
\address{$^3$Instituto de Ciencias Matem\'{a}ticas (CSIC - UAM - UC3M - UCM), Nicol\'{a}s Cabrera 13-15, Campus de Cantoblanco, UAM, 28049, Madrid, Spain.}


\ead{agarcia@math.uc3m.es, miguelangel.hernandez.medina@upm.es, albertoi@math.uc3m.es}



\begin{abstract}
Nyquist-Shannon sampling theorem, instrumental in classical telecommunication technologies, is extended to quantum systems supporting a unitary representation of a finite group $G$.   Two main ideas from the classical theory having natural counterparts in the quantum setting: frames and invariant subspaces, provide the mathematical background for the theory.    The main ingredients of classical sampling theorems are discussed and their quantum counterparts are thoroughly analyzed in this simple situation.   A few examples illustrating the obtained results are discussed. 
\end{abstract}


\vspace{2pc}


\submitto{\JPA}

\noindent{\bf AMS}: 20C40; 42C15; 94A20.

\maketitle

\section{Introduction}
\label{section1}

Shannon's celebrated theorems:  Shannon-Hartley's channel capacity theorem \cite{Ha28,Sh49}, Shannon's source coding  theorem  \cite{Sh48} and Nyquist-Shannon sampling theorem \cite{Ny28,Sh49},  constitute the backbone of the mathematical background of modern telecommunication technologies.  The development of Quantum Information Technologies have caused that the first two theorems have already been extended to the quantum setting (see for instance \cite{De08}, \cite{Wi13} and references therein) and their quantum counterparts are today part of the mainstream of Quantum Information.

Interestingly enough, Nyquist-Shannon sampling theorem has not found a place in Quantum Information yet.   
Actually, as far as we know, there is not a genuine quantum version of it.  The classical sampling theorem states, in its more streamlined form, that band-limited signals $f(t)$ (within the interval $[-\pi,\pi]$ for instance) can be reconstructed from the family of samples $\{f(k)\}$ and a given family of signals $S_k(t)$, $k$ being an integer. More precisely:
 \begin{equation}
 \label{shannon}
 f(t) = \sum_{k\in \ZZ} f(k) \frac{\sin \pi (t-k)}{\pi (t-k)} \,, \qquad t\in \RR\,.
 \end{equation}  
 
 It is obvious the interest that sampling formulas, similar to the previous one, would have in the quantum setting as they would provide an alternative way to describe the states of quantum systems.   Thus, if we consider, instead of signals, vectors $|x\rangle$ in a Hilbert space describing pure states of a quantum system, a ``quantum sampling theorem'' would provide a way of reconstructing the states from a family of ``samples'', denoted in what follows by $\{\mathcal{L}x(k)\}$, $k$ an index, and a prescribed set of quantum states $\{|C_k\rangle\}$, the analogues of the signals $\big\{S_k(t) =  \frac{\sin \pi (t-k)}{\pi (t-k)}\big\}$ in classical sampling theory, Eq. (\ref{shannon}), as $|x\rangle = \sum_k \mathcal{L}x(k) |C_k\rangle$.  Such auxiliary states could be manufactured independently of the system under scrutiny and all that would be needed will be to prepare a superposition of them with coefficients the previously obtained samples.  
 
 Arguably, it can be stated that any orthonormal basis $\{|e_k\rangle\}$ provides a sampling theorem with samples the amplitudes $\mathcal{L}x(k) = \langle e_k | x\rangle$ and with auxiliary states the orthonormal vectors themselves, i.e., $| C_k \rangle = | e_k \rangle$.    Actually, Shannon's formula (\ref{shannon}) is just the orthonormal expansion in the Hilbert space of square integrable functions in the real line of the function $f$ with respect to the orthonormal basis $\{S_k\}$.   Notice however that the orthonormal basis given by the shifted sine-cardinal functions $S_k (t) = \frac{\sin \pi (t-k)}{\pi (t-k)}$ has the additional property that the coefficients of the expansion are the actual values of the function $f$ itself, and the word ``sample'' has, in this context, the meaning of the actual value of the function     which is precisely the fact that provides the ground for its technological implementation and it is one of the main properties of Shannon's theorem that we would like to extend to the quantum setting.
   
We must point out that, certainly, quantum tomography (see for instance \cite{As15},  \cite{Ib09} and references therein) aims also in the direction sketched above, however the current status of the theory doesn't allow for a clear cut reconstruction theorem as the ones provides by Shannon's theorem and its generalizations.   

Thus the aim of the present paper is to walk the first steps into the construction of a rigorous and broad enough sampling theory for quantum systems inspired directly in the classical sampling theorem by Nyquist-Shannon.
In order to achieve it, we will use some recent ideas underlying a family of classical generalized 
Shannon's theorems.  Two ingredients appearing in these generalizations seem to be critical for the development of a quantum version: the use of frames and shift-invariant subspaces \cite{garcia:06}. 

\bigskip 

Frames provide a convenient and flexible tool extending the notion of orthonormal or Riesz  bases in Hilbert spaces \cite{ole:03}. Traditionally, frames were used in signal and image processing, non-harmonic analysis, data compression, and sampling theory, but nowadays
frame theory plays also a fundamental role in a wide variety of problems in both pure  and applied  mathematics, computer science, physics and engineering.  Let us recall that a frame is any sequence of vectors $\{| c_k \rangle\}$ in a Hilbert space such that there exist two constants $0< A \le B$ such that for any vector $| x \rangle$ we have 
\[
A || x ||^2 \le \sum_k | \langle c_k \mid x \rangle |^2 \le B || x ||^2\,. 
\]

Given a frame $\{| c_k \rangle\}$ in a Hilbert space the representation 
property of any vector $| x \rangle$ as a series $\bra{x}=\sum_k \alpha_k \bra{c_k}$ is retained, 
but, unlike the case of orthonormal (Riesz) bases, the uniqueness 
of this representation (for overcomplete frames) is sacrificed. 
Suitable frame coefficients 
$\alpha_k$ which depend continuously and linearly on $\bra{x}$ 
are obtained by using dual frames $\{\bra{d_k}\}$ of $\{\bra{c_k}\}$, i.e., 
$\{\bra{d_k}\}$ is another frame for the Hilbert space such that 
$\bra{x}=\sum_k \langle d_k \mid x \rangle \,\bra{c_k}=\sum_k \langle c_k \mid x \rangle \,\bra{d_k}$. Recall that a  Riesz basis in a separable Hilbert space is the image of an orthonormal basis by means of a bounded invertible operator; a Riesz basis has a unique dual (biorthogonal) basis. A Riesz sequence in a Hilbert space is a Riesz basis for its closed span.
For more details on the theory of frames see, for instance, the monograph \cite{ole:03} and references therein.
   
The redundancy of frames, which gives flexibility and robustness, is the key to their significance for applications (see, for instance, the nice introduction in Chapter1 of Ref.~\cite{casazza:14} and references therein).   It is worth to note that frames in finite dimension are nothing but spanning sets of vectors. Frames have already been considered in  problems related to this work (see for instance \cite{An98}, \cite{An08}, or more recently \cite{Ba15}). 

\bigskip  

The second ingredient consists of a natural extension of the notion of shift-invariant subspaces of $L^2(\RR)$, that is, subspaces of functions invariant under the shift transformation $f(t) \mapsto f(t-1)$, that play a fundamental role in classical sampling theory \cite{garcia:06} (see also \cite{aldroubi:01,sun:03}).  The shift transformation defines a unitary operator on the Hilbert space of square integrable functions on the real line.  Hence it is natural to consider subspaces of a Hilbert space which are invariant under a given unitary transformation $U$ and to extend to this situation the results of classical sampling theory.  This idea has been recently considered in a number of papers (see for instance the related works \cite{hector:13,hector:14,hector:15,garcia:15} and \cite{michaeli:11,pohl:12}). 

\medskip  

More concretely, to visualize better the role of these two ingredients and to prepare the ground for the sampling theorem to be derived in this paper, let us describe the generalized classical sampling theorem that can be obtained combining them and that we want to extend to the quantum setting.  Indeed, let $V_{{\rm samp}}$ be the Hilbert space where we want to obtain a sampling formula involving the sequence $\{m_k(x)\}$ of measurements (samples) obtained from each $\bra{x}\in V_{{\rm samp}}$.  The general mathematical procedure can be summarized as follows (see \cite{garcia:00,garcia:06,hector:14} for details):
\begin{enumerate}
\item Express the available samples $\{m_k(x)\}$ of $\bra{x}\in V_{{\rm samp}}$ as the inner products $m_k(x)=\langle f_k\mid f\rangle_{L^2}$ in an auxiliary $L^2$-Hilbert space,  where $\{f_k\}$  is  some fixed sequence in $L^2$, and  $\bra{x}$ is the image under an isomorphism $\Tc_V: L^2 \rightarrow V_{{\rm samp}}$ of the function $f$.
\item Characterize the above sequence $\{f_k\}$ as a frame for the $L^2$-auxiliary space.
\item Find the dual frames $\{g_k\}$ of $\{f_k\}$. Thus we have
\[
f=\sum_k \langle f_k \mid f\rangle \, g_k=\sum_k m_k(x)\, g_k\,.
\]
\item Finally, applying the isomorphism $\Tc_V$ in the above expansion we get a sampling formula in $V_{{\rm samp}}$. Namely,
\[
\bra{x}=\sum_k m_k(x)\, \bra{\Tc_V(g_k)}\,.
\]
Moreover, due to the unitary invariant character of $V_{{\rm samp}}$, the isomorphism $\Tc_V$ satisfies a {\em shifting property} which simplifies the obtention of the reconstruction vectors $\bra{\Tc_V(g_k)}$.
\end{enumerate}
In  the particular case of the classical sampling theorem, Eq. (\ref{shannon}), we have that any band-limited function $f$ can be expressed as 
\begin{equation}
\label{expression}
f(t)=\big\langle \frac{1}{\sqrt{2\pi} } {\rm e}^{-i\pi wt} | \widehat{f}\, \big\rangle_{L^2[-\pi, \pi]}\,, \quad t\in \RR\,, 
\end{equation}
where $\widehat{f}$ denotes the Fourier transform of $f$. In particular, for the samples of $f$ at $k \in \ZZ$ we have that  $f(k)=\big\langle {\rm e}^{-i\pi kt}/\sqrt{2\pi} \mid \widehat{f}\, \big\rangle_{L^2[-\pi, \pi]}$, and the isomorphism $\Tc_V$ is the inverse Fourier transform $\mathcal{F}^{-1}$. Since the sequence $\big\{{\rm e}^{-i\pi kt}/\sqrt{2\pi}\big\}_{k\in \ZZ}$ is an orthonormal basis for $L^2[-\pi, \pi]$ we have
\[
\widehat{f}(w)=\sum_{k\in \ZZ} f(k)\, \frac{{\rm e}^{-i\pi kw}}{\sqrt{2\pi}} \qquad \mathrm{in} \, \, L^2[-\pi, \pi]\,,
\]
and applying $\mathcal{F}^{-1}$,
\[
f(t)=\sum_{k\in \ZZ} f(k)\,\mathcal{F}^{-1}\big( \frac{{\rm e}^{-i\pi kw}}{\sqrt{2\pi}}\chi_{[-\pi, \pi]}(w)\big)(t)=\sum_{k\in \ZZ} f(k) \frac{\sin \pi (t-k)}{\pi (t-k)}\quad \mathrm{in} \, \, L^2(\RR) \,.
\]
Notice that we have used that $\mathcal{F}^{-1}\big(\chi_{[-\pi, \pi]}(w)\big)(t)=\frac{\sin \pi t}{\pi t}$ and the shifting property which satisfies the Fourier transform. Cauchy-Schwarz's inequality in (\ref{expression}) proves the inequality
$|f(t)|\le \|f\|$, $t\in \RR$. In other words, here the convergence in norm implies pointwise convergence which is uniform on $\RR$.

\medskip

Classical sampling has been generalized in the following way: In $L^2(\mathbb{R})$ consider the shift operator $U: f(t) \mapsto f(t-1)$. The functions (signals) to be sampled belong to some (principal) shift-invariant subspace $V_\varphi^2:= \overline{{\rm span}}_{L^2(\mathbb{R})}\big\{\varphi(t-n),\;  n\in \mathbb{Z}\big\}$, where the generator function $\varphi$ belongs to $L^2(\mathbb{R})$ and the sequence $\{\varphi(t -n)\}_{n\in \mathbb{Z}}$ is a Riesz sequence for $L^2(\mathbb{R})$. Thus, the shift-invariant space $V_\varphi^2$ can be described as
\[
V_\varphi^2=\Big\{\sum_{n \in \mathbb{Z}} \alpha_{n} \ \varphi(t-n)\,:\, \{\alpha_n\}_{n\in \mathbb{Z}} \in \ell^2(\mathbb{Z})\Big\} 
\]
In particular, spline or wavelet spaces are examples of shift-invariant subspaces $V_\varphi^2$.

On the other hand, in many common situations the available data are samples of some filtered versions $f*\mathsf{h}_j$ of the signal $f$ itself, where the average function $\mathsf{h}_j$ reflects the characteristics of the adquisition device. For $N$ convolution systems (linear time-invariant systems or filters in engineering jargon) $\mathcal{L}_jf:=f*\mathsf{h}_j$,\ $j=1,2,\ldots,N$, assume that, for any $f$ in $V_\varphi^2$, the sequence of samples $\{(\mathcal{L}_j f)(rm)\}_{m\in \mathbb{Z};\,j=1,2,\ldots,N}$ is available, where $r\in \mathbb{N}$ denotes the sampling period. The generalized sampling problem mathematically consists of the stable recovery of any $f\in V_\varphi^2$ from the above sequence of its samples. In other words, it deals with the construction of sampling formulas in $V_\varphi^2$ having the form
\begin{equation}
\label{general_sampling}
f(t)=\sum_{j=1}^N\sum_{m\in\mathbb{Z}}\big(\mathcal{L}_jf\big)(rm)\, S_j(t-rm)\,, \qquad t\in \mathbb{R}\,,
\end{equation}
where the sequence of reconstruction functions $\{S_j(\cdot-rm)\}_{m\in \mathbb{Z};\, j=1,2,\ldots,N}$ is a frame for the shift-invariant space $V_\varphi^2$. 
In this setting, the isomorphism $\Tc_{V_\varphi^2}: L^2(0, 1) \rightarrow V_\varphi^2$ maps the orthonormal basis $\{{\rm e}^{-2\pi i nw}\}_{n\in \ZZ}$ onto $\{\varphi(t -n)\}_{n\in \ZZ}$, and it satisfies the shifting property $\mathcal{T}_{V_\varphi^2}(e^{-2\pi inx}F) = (\mathcal{T}_{V_\varphi^2} F)(t- n)$ for $F\in L^2(0, 1)$, $n \in\mathbb{Z}$. Similar results can be obtained if we consider a unitary operator $U$ in an abstract Hilbert space instead of the shift operator in $L^2(\RR)$ (see \cite{hector:13,hector:14,pohl:12}).

\bigskip

In this paper we will consider an extension of the previous generalized sampling formula, Eq.~(\ref{general_sampling}), to the realm of an abstract  complex separable Hilbert space $\mathcal{H}$ whose rays represent the pure states of a quantum system.  The notion of $U$-invariant subspaces will be extracted from a situation which is common in physical systems and that corresponds to the presence of a group represented unitarily on the state space of the system.   That is, we will consider a group $G$ and a unitary representation $U \colon G \to \mathcal{U}(\mathcal{H})$.   Hence, the main notion that will be used henceforth will be that of subspaces invariant under the representation of the group $G$, i.e., closed subspaces $W \subset \mathcal{H}$ such that $U(g) W \subset W$ for all $g\in G$.  Clearly the notion of $U$-invariant subspace corresponds to the choice of the Abelian group of integers $\mathbb{Z}$ and the unitary representation provided by $U(n) = U^n$ where $U$ is a given unitary operator.    The theory developed in what follows can also be considered as a non-commutative extension of the standard classical sampling theory once we restrict ourselves to Hilbert spaces which are Hilbert spaces of functions defined on a suitable measure space.   

Only finite groups $G$ will be discussed here, both for simplicity of exposition and because many relevant physical systems exhibit them in a natural way as symmetry groups: from chemistry and molecular physics  \cite{Bu05,Ke07,Wi59}, to  condensed matter physics \cite{Dr08}.    It is worth to point it out that it is not necessary that the group $G$ would be a symmetry group of the system.   Actually, if the group is a symmetry group, we will get and additional information on the form of a sampling formula for the time-evolution of the given state, however, in order to construct the sampling expansions exhibited below, is unnecessary.  

The use of frames, finite under the circumstances, will be natural within this context.    The auxiliary space used in the theory will be the space of square integrable functions in the group, that in this context agrees with the group algebra of the group $G$.

The first part of the paper, Section \ref{section2}, will be devoted to establish the mathematical setting of the problem, to discuss the main ideas used later on, like the notion frames. and to introduce the main notion of generalized samples and their properties.   In Section \ref{section3} the main theorem stating the sampling formula for a family of states of quantum systems supporting an unitary representation of a finite group will be established.
Finally, in Section \ref{section4} some illustrative examples will be discussed, in particular the simple cases of cyclic and dihedral groups which are fundamental in the applications of the theory.

\section{The mathematical setting}
\label{section2}
Let $G$ be a finite (not necessarily commutative) group with identity element $e$; we denote its order as $|G|$. Let also $g\in G \mapsto U(g)\in \Uc(\Hc)$ be a unitary representation of  a  $G$ in a complex separable Hilbert space $\Hc$, i.e., a homomorphism from the groups $G$ into the group $\Uc(\Hc)$ of unitary operators on $\Hc$, i.e., a map satisfying  $U(gg')=U(g)U(g')$ and $U(e)=I_{\Hc}$. 

Let $W$ be a closed subspace of $\mathcal{H}$ invariant under the action of the group $G$ provided by the representation $U$, that is $U(g) |x\rangle \in W$ for every $|x\rangle \in W$ and $g \in G$.   Clearly the Hilbert space $W$ supports a unitary representation of the group $G$ too.  Moreover because the group $G$ is finite\footnote{Similar considerations would apply for a much larger class of groups.} any unitary representation is completely reducible into irreducible components.    

However, for reasons that will be clear along the text, it will be better to decompose the Hilbert space in cyclic invariant subspaces, that is, invariant subspaces $W$ possessing a cyclic vector $|a\rangle$, hence $W = \mathrm{span}\{ U(g)|a\rangle \mid g \in G \}$.   We will denote such space by $\mathcal{A}_a$ and it is clear that $\mathcal{H}$ can be decomposed as a countable direct sum of mutually orthogonal such subspaces $\mathcal{H} = \bigoplus_{n = 1}^\infty W_{n}$, $W_n = \mathcal{A}_{a_n}$, for a family of vectors $a_n\in \mathcal{H}$.  Consequently the problem of sampling a given state $|x\rangle \in \mathcal{H}$ is reduced to the problem of sampling the components $\bra{x_n} \in \mathcal{A}_{a_n}$ such that $\bra{x} = \sum_n \bra{x_n}$.   
 
\medskip

So, from now on, we will consider a fixed vector $\bra{a}\in  \Hc$ and the subspace $\Ac_a$  of $\Hc$ spanned by $\bra{a_g}  = U(g)\bra a$,  $g\in G$. In case this set is linearly independent in $\Hc$, each $\bra{x}\in \Ac_a$ can be expressed as the unique expansion $\bra{x}= \sum_{g\in G} \alpha_g   \,U(g)\bra{a}  $, with $\alpha_g  \in \CC$.

There is a close relationship between the (finite) sequence $\bra{a_g} $ and the so-called {\em stationary sequences} (see Kolmogorov \cite{kolmogorov:41}). We say that the (finite) sequence $\{\bra{x_g}    \,:\, g\in G\}$ in $\Hc$ is (left) {\em $G$-stationary} if:
\[
\langle{x_g} \bra{x_{g'}}_\Hc=\langle x_{hg} | x_{hg'} \rangle_\Hc\,,\quad \mathrm{for all} g, g', h\in G \,.
\]
Then it is easy to deduce that  there exists a unitary representation $U(g)$ of $G$ and an $\bra{a}\in  \Hc$ such that $\bra{x_g}   =U(g)\bra{a}  $, $g\in G$. Thus we define the {\em auto-covariance} of the finite sequence $\big\{U(g)\bra{a}  \big\}_{g\in G}$ as
\[
R_a(g):=\langle a | U(g) |a\rangle_\Hc\,, \quad g\in G\,.
\] 
Similarly, we can define the {\em cross-covariance} between the finite sequences $\big\{U(g)\bra{a}  \big\}_{g\in G}$ and $\big\{U(g)\bra b\big\}_{g\in G}$ where $\bra a, \bra b \in\Hc$ as
\[
R_{a,b}(g):=\langle a | U(g) | b\rangle_\Hc\,, \quad g\in G\,.
\]
Note that $R_{a,b}(g)=\overline{R_{b,a}(g^{-1})}$ for $\bra a, \bra b \in\Hc$ and $g\in G$.

\medskip

\begin{prop}
Let $\mathbf{R}_a$ denote the $|G|\times |G|$ square matrix where $|G|$ is the order of the group, defined by $\mathbf{R}_a:=\big(R_a(h^{-1}g) \big)_{(h,g)\in G\times G}$. Then, the set of vectors $\big\{U(g)\bra{a}  \,:\, g\in G\big\}$ is linearly independent in $\Hc$ if and only if $\det \mathbf{R}_a \neq 0$.
\end{prop}

\noindent\textit{Proof:}
If $\det \mathbf{R}_a = 0$ then there exists a vector $\boldsymbol{\lambda}=(\lambda_g )_{g\in G} \in \CC^{|G|}$ such that  $\boldsymbol{\lambda} \neq \mathbf{0}$ and  $\mathbf{R}_a  \boldsymbol{\lambda} = 0$. Thus  $\sum_{g\in G} \lambda_g \,U(g)\bra{a}  $ is orthogonal 
to $U(g)\bra{a}  $ for all $g\in G$ so that  $\sum_{g} \lambda_g \,U(g)\bra{a}  =0$.  Conversely, if $\sum_{g\in G} \lambda_g \,U(g)\bra{a}  =0$ for some $\boldsymbol{\lambda} \neq \mathbf{0}$ then the inner product  in the above expression  with each $U(h)\bra{a} $, $h\in G$, yields   
$\mathbf{R}_a  \boldsymbol{\lambda} = 0$. \hfill$\Box$
\subsection*{The isomorphism $\Tc^G_a$}

Consider the group algebra $\mathbb{C}[G]$, that is, the complex linear space generated by the elements of the group $G$.  Thus $\mathbb{C}[G]$ has dimension $|G|$ and its elements can be identified with the space of functions $ \boldsymbol{\alpha} \colon G \to \mathbb{C}$, $g \mapsto \boldsymbol{\alpha}(g)$; in brief $\boldsymbol{\alpha}=\big( \boldsymbol{\alpha}(g)\big)_{g\in G}$.   In the case we are dealing here of finite groups, such functions are all obviously integrable and square integrable, hence it can be identified with $L^2(G)$ that endowed with its natural inner product $\langle \boldsymbol{\alpha}\mid \boldsymbol{\beta} \rangle$ becomes a Hilbert space isomorphic to $\CC^{|G|}$. 

The Hilbert space $L^2(G)$ supports a natural unitary representations of $G$ called the {\em left regular representation} $L_s$, $s\in G$, defined by (similarly on the right):
\[
L_s \boldsymbol{\alpha}(g)=\boldsymbol{\alpha}(s^{-1}g)\quad \mathrm{for} \, \, s,g\in G\,.
\]
Next we define the following isomorphism $\Tc^G_a$, analogous to the isomorphism $\mathcal{T}_V$ used in classical sampling theory discussed in the introduction,  between $L^2(G)$ and $\Ac_a$:
\begin{equation}
\label{iso}
\begin{array}[c]{ccll}
\Tc^G_a: & L^2(G) & \longrightarrow & \mathcal{A}_a\\
       & \boldsymbol{\alpha} & \longmapsto & \bra{x}= \displaystyle{\sum_{g\in G} \boldsymbol{\alpha}(g)\,U(g)\bra{a}  }\,.
\end{array}
\end{equation}
This isomorphism $\Tc^G_a$ has the following {\em shifting property} with respect to the left regular representation $L_s$:
\begin{prop}
For any $s\in G$ and $\boldsymbol{\alpha}\in L^2(G)$ we have that
\begin{equation}\label{shiftgroup}
\Tc^G_a\big(L_{s} \boldsymbol{\alpha}\big)=U(s)\,\Tc^G_a (\boldsymbol{\alpha})
\end{equation}
\end{prop}
\noindent\textit{Proof:}
Indeed, denoting $g'= s^{-1}g$ we have
\begin{eqnarray}
\Tc^G_a\big(L_{s} \boldsymbol{\alpha}\big) &=& \sum_{g\in G} \boldsymbol{\alpha}(s^{-1}g) U(g)\bra{a}  =\sum_{g'\in G} \boldsymbol{\alpha}(g') U(sg')\bra{a} \\
&=&\sum_{g'\in G} \boldsymbol{\alpha}(g') U(s)U(g')\bra{a}=U(s)\,\Tc^G_a (\boldsymbol{\alpha})
\end{eqnarray}
\hfill$\Box$

Notice that the shifting property (\ref{shiftgroup}) is just the alluded shifting property in the introduction which is satisfied by the isomorphism $\mathcal{T}_V$. 
\subsection*{An expression for the generalized samples}

For our sampling purposes we consider here an Abelian subgroup $K$ of $G$ (not necessarily normal) such that it possesses a {\em complement} $H$, i.e., $H$ is a subgroup of $G$ such that $KH=G$ and $K\cap H=\{e\}$ (in particular, $K$ is a {\em transversal} of $H$).    

In case that $K$ is a normal subgroup, the Schur-Zassenhaus theorem gives us a sufficient condition for the existence of a complement of $K$ in $G$. Namely, if $K$ is an Abelian normal subgroup of $G$ such that $|K|$ and $|G/K|$ are coprime then there exists a complement $H$ of $K$ in $G$.   For more details see, for instance, \cite[p.76]{isaacs:08}.    In such case the group $G$ is the {\em semidirect product} of $K$ and $H$, and denoted as $G = H\rtimes K$. 

Moreover if $G$ is the semidirect product of the Abelian subgroup $K$ (not necessarily normal) and the normal subgroup $H$, i.e., $G = K \rtimes H$, then $G = KH$, $K\cap H=\{e\}$ and $H$ is a complement of $K$.   This is the situation which is commonly found in applications to specific physical systems.

As a consequence we can choose an element  
 of $K$ in each coset of the quotient set $G/H$. Thus, denoting $\ell:=|K|=|G|/|H|$ in case $K=\big\{\tau_0=e, \tau_1, \cdots, \tau_{\ell-1}\big\}$ we can describe the quotient set $G/H$ as 
\[
G/H=\big\{[e=\tau_0], [\tau_1], \cdots, [\tau_{\ell-1}] \big\}\,.
\]
From now on we write the group $G$ as (for any fixed way of writing the elements of $H$) 
\[
G=\big\{\tau_0^{-1}H, \tau_1^{-1}H, \cdots, \tau_{\ell-1}^{-1}H \big\}\,.
\]
Fixed $N$ elements $b_j\in\Hc$, $j=1,2,\dots,N$, for each $\bra{x}= \sum_{s\in G}\alpha_s \,U(s)\bra a \in \Ac_a$ we define its {\em generalized samples} by
\begin{equation}
\label{gsamples}
\Lc_j x (\tau_n):=\big\langle U(\tau_n)b_j \mid x\big \rangle_\Hc\,, \quad n=0, 1, \dots, \ell-1 \mathrm{and} \, \, j=1, 2, \dots,N \,.
\end{equation}
We will refer to each $\Lc_j$ as a $\Lc_j$-system acting on $\Ac_a$, $j=1, 2, \dots,N$.  

Notice that the expression for the generalized samples (\ref{gsamples}) is an straightforward generalization of the convolution of the sampled vector $\bra{x}$ with the vectors $\bra{b_j}$.

Besides, to recover any $\bra{x} \in \Ac_a$ we need at least $|G|$ samples; if we are sampling at $K$, we will need at least $N$ $\Lc_j$-systems such that $N\ell \geq |G|=\ell |H|$, i.e., $|N|\geq |H|$.

As it was discussed in the introduction, the main goal of this paper is to recover any state vector $\bra{x}\in    \Ac_a$ by means of its generalized samples (\ref{gsamples}) and sampling formulas taking care of the unitary structure of $\Ac_a$. To this end, we first obtain an alternative expression for the generalized sample $\Lc_jx(\tau_n)$ with $n=0, 1, \dots, \ell-1$. Namely,

\begin{eqnarray}
\Lc_jx(\tau_n)&=\big\langle U(\tau_n)b_j \mid x\big \rangle_\Hc=\Big\langle U(\tau_n)b_j \mid \sum_{s\in G} \alpha_s \,U(s)a\Big\rangle_\Hc  \nonumber \\
         &=\sum_{s\in G} \alpha_s \,\big\langle U(\tau_n)b_j \mid U (s)a\big\rangle_\Hc 
         =\big\langle G_{j,\tau_n}\mid \boldsymbol{\alpha}\big\rangle_{L^2(G)}\,, \label{genSamples}
\end{eqnarray}
where $\boldsymbol{\alpha}=(\alpha_s)_{s\in G}$ and $G_{j,\tau_n}=(\overline{\langle U(\tau_n)b_j \mid U(s)a \rangle})_{s\in G}$ belong to $L^2(G)$. 

The vectors $G_{j,\tau_n}\in L^2(G)$, $j=1,2,\dots,N$, $n=0, 1, \dots, \ell-1$, can be expressed, in terms of the  cross-covariances $R_{b_j,a}$,  as 

\begin{eqnarray*}
  G_{j,\tau_n}&=\big(\langle U(s)a \mid U(\tau_n)b_j \rangle\big)_{s\in G}=
         \big(\langle a \mid U(s^{-1}\tau_n) b_j \rangle\big)_{s\in G} \\
         &= \big(R_{a,b_j}(s^{-1}\tau_n)\big)_{s\in G}= \big(\overline{R_{b_j,a}(\tau_n^{-1}s})\big)_{s\in G}\,.
\end{eqnarray*}

Having in mind expression (\ref{genSamples}) for the samples and the isomorphism $\Tc_a^G$ defined in (\ref{iso}) we deduce the following result (see also the finite frame theory in \cite{casazza:14}):
\begin{prop}
\label{prop3}
Any $\bra{x}\in     \Ac_a$ can be recovered from its samples $\big\{\Lc_j x (\tau_n) \big\}$, $j=1,2,\ldots, N$, $n=0,1,\ldots,\ell-1$ if and only if the set of vectors $\{G_{j,\tau_n}\}$, $j=1,2,\ldots, N$, $n=0,1,\ldots,\ell-1$, in $L^2(G)$ form a frame (a spanning set) for $L^2(G)$. 
\end{prop}
Equivalently, the $|G|\times N\ell$ matrix
\begin{equation}
\label{matrixG}
\left(\begin{array}{cccccccccc}
\vdots & \vdots & \vdots & \vdots & \vdots & \vdots & \vdots & \vdots & \vdots& \vdots\\
G_{1,\tau_0} & \cdots & G_{1,\tau_{\ell-1}} & G_{2,\tau_0} & \cdots &G_{2,\tau_{\ell-1}} 
      & \cdots & G_{N,\tau_0} & \cdots &G_{N,\tau_{\ell-1}}\\
\vdots & \vdots & \vdots & \vdots & \vdots & \vdots & \vdots & \vdots & \vdots &\vdots
\end{array}\right)
\end{equation} 
has rank $|G|$. Hence, we have that $|G|\leq N \ell$, that is, the number of needed $\Lc_j$-systems is necessarily $N\geq |H|$. The vectors $G_{j,\tau_n}$ can be written as column matrices as
\[
G_{j,\tau_n}=\Big(\overline{R_{b_j,a}}(\tau^{-1}_{n}\tau_0^{-1} H), 
              \overline{R{b_j,a}}(\tau^{-1}_{n}\tau_1^{-1} H),\dots, \overline{R{b_j,a}}(\tau^{-1}_{n}\tau_{\ell-1}^{-1} H)\Big)^\top
\]
where $\tau_n\in K$ and $\overline{R_{b_j,a}}(\tau^{-1}_{n}\tau_p^{-1} H)$ is given by
\[
  \overline{R_{b_j,a}}(\tau^{-1}_{n}\tau^{-1}_p H)=\Big(\overline{R_{b_j,a}}(\tau^{-1}_{n}g_{p_1}), 
                        \overline{R_{b_j,a}}(\tau^{-1}_{n}g_{p_2}), \dots, \overline{R_{b_j,a}}(\tau^{-1}_{n}g_{p_{|H|}})\Big)
\]
being $\tau_p^{-1}H=\{g_{p_1},\dots,g_{p_{|H|}}\}$ with $p=0, 1, \dots, \ell-1$. 

\medskip

For each $j=1,2,\dots,N$ let $\mathbf{R}_{b_j,a}$ be the $\ell\times |G|$ matrix 
\[
  \mathbf{R}_{b_j,a}=
      \left(\begin{array}{cccc}
          R_{b_j,a}(\tau^{-1}_{0}\tau_0^{-1} H)&
              R_{b_j,a}(\tau^{-1}_{0}\tau_1^{-1} H)&\dots& R_{b_j,a}(\tau^{-1}_{0}\tau_{\ell-1}^{-1} H)\\
        R_{b_j,a}(\tau^{-1}_{1}\tau_0^{-1} H)&
              R_{b_j,a}(\tau^{-1}_{1}\tau_1^{-1} H)&\dots& R_{b_j,a}(\tau^{-1}_{1}\tau_{\ell-1}^{-1} H)\\
          \vdots & \vdots & \cdots& \vdots\\
         R_{b_j,a}(\tau^{-1}_{\ell-1}\tau_0^{-1} H)&
              R_{b_j,a}(\tau^{-1}_{\ell-1}\tau_1^{-1} H)&\dots& R_{b_j,a}(\tau^{-1}_{\ell-1}\tau_{\ell-1}^{-1} H) 
         \end{array}\right)
\]
Since $K$ is an Abelian subgroup of $G$ the cosets $\tau^{-1}_n\tau_p^{-1} H$ and $\tau^{-1}_p\tau_n^{-1} H$
coincide. As a consequence, $\mathbf{R}_{b_j,a}$ is the block symmetric matrix
\begin{equation}
 \label{estructuraRCasoGeneral}
  \mathbf{R}_{b_j,a}=
     \left( \begin{array}{cccc}
          R_{b_j,a}(\tau^{-1}_{0}\tau_0^{-1} H)&
              R_{b_j,a}(\tau_1^{-1}\tau^{-1}_{0} H)&\dots& R_{b_j,a}(\tau_{\ell-1}^{-1}\tau^{-1}_{0} H)\\
        R_{b_j,a}(\tau_0^{-1}\tau^{-1}_{1} H)&
              R_{b_j,a}(\tau^{-1}_{1}\tau_1^{-1} H)&\dots& R_{b_j,a}(\tau_{\ell-1}^{-1}\tau^{-1}_{1} H)\\
          \vdots & \vdots & \cdots& \vdots\\
         R_{b_j,a}(\tau_0^{-1}\tau^{-1}_{\ell-1} H)&
              R_{b_j,a}(\tau_1^{-1}\tau^{-1}_{\ell-1} H)&\dots& R_{b_j,a}(\tau^{-1}_{\ell-1}\tau_{\ell-1}^{-1} H) 
      \end{array} \right)
\end{equation}
The matrix given in (\ref{matrixG}) can be written as $\big(\mathbf{R}_{b_1,a}^*\,\mathbf{R}_{b_2,a}^*\,\dots \,\mathbf{R}_{b_N,a}^*\big)$, where the symbol $*$ denotes the traspose conjugate matrix. Thus, Proposition \ref{prop3} can be restated in terms of the $N\ell\times |G|$ matrix of cross-covariances
$\mathbf{R}_{\mathbf{b},a}$  defined by
\begin{equation}
\label{matrixcc}
\mathbf{R}_{\mathbf{b},a}:= 
\left( \begin{array}{c}       
\mathbf{R}_{b_1,a} \\
\mathbf{R}_{b_2,a}\\  
\vdots \\
\mathbf{R}_{b_N,a}   \\
\end{array}\right)\,.
\end{equation}
\begin{cor}
\label{cor1}
Any $\bra{x}\in     \Ac_a$ can be recovered from its samples $\big\{\Lc_j x (\tau_n) \big\}$, $j=1,2,\ldots, N$, $n=0,1,\ldots,\ell-1$ if and only $\rank\, \mathbf{R}_{\mathbf{b},a}=|G|$. 
\end{cor}
Besides, Equation (\ref{genSamples}) can be expressed, for any $\bra{x}= \sum_{s\in G}\alpha_s \,U(s)\bra a$ in $\Ac_a$,  as
\[
   \left(\begin{array}{c}
     \Lc_j x(\tau_0)\\
     \Lc_j x(\tau_1)\\
          \vdots\\
     \Lc_j x(\tau_{\ell-1})
   \end{array}\right) = \mathbf{R}_{b_j,a}\,\boldsymbol{\alpha}
\]
where $\boldsymbol{\alpha}=(\alpha_s)_{s\in G}$. As a consequence we deduce the samples expression:
\begin{prop} 
For any $\bra{x}= \sum_{s\in G}\alpha_s \, U(s)\bra{a}$   in $\Ac_a$ consider its samples vector 
\begin{equation}
\Lc_{\mathrm{samp}}x= \big(\Lc_1 x(\tau_0)\dots\Lc_1 x(\tau_{\ell-1})\cdots \Lc_N x(\tau_0)\dots \Lc_N x(\tau_{\ell-1})\big)^\top\,.
\end{equation}
Then, the matrix relationship
\begin{equation}
\label{samplematrix}
\Lc_{\mathrm{samp}}x= \mathbf{R}_{a,\mathbf{b}}\,\boldsymbol{\alpha}
\end{equation}
holds, where $\boldsymbol{\alpha}=(\alpha_s)_{s\in G}$ and $\mathbf{R}_{\mathbf{b},a}$ is the $N\ell\times |G|$ matrix of cross-covariances defined in (\ref{matrixcc}).
\end{prop}

\medskip
Assuming that $\{G_{j,\tau_n}\}$, $j=1,2,\ldots, N$, $n=0,1,\ldots,\ell-1$, is a frame for $L^2(G)$ we have that the rank of the matrix  $\mathbf{R}_{\mathbf{b},a}$ is $|G|$. Thus the  Moore-Penrose  pseudoinverse of $\mathbf{R}_{\mathbf{b},a}$ is the $|G|\times N\ell$ matrix  
$\mathbf{R}^+_{\mathbf{b},a}=[\mathbf{R}^*_{\mathbf{b},a}\mathbf{R}_{\mathbf{b},a}]^{-1}\mathbf{R}^*_{\mathbf{b},a}$ (see \cite{penrose:55}). Writing the columns of $\mathbf{R}^+_{\mathbf{b},a}$ as
\[
\mathbf{R}^+_{\mathbf{b},a}=
\left( \begin{array}{cccccccccc}
\vdots & \vdots & \vdots & \vdots & \vdots & \vdots & \vdots & \vdots & \vdots& \vdots\\
\mathbf{R}^+_{1,\tau_0} & \cdots & \mathbf{R}^+_{1,\tau_{\ell-1}} & \mathbf{R}^+_{2,\tau_0} & \cdots &\mathbf{R}^+_{2,\tau_{\ell-1}} 
      & \cdots & \mathbf{R}^+_{N,\tau_0} & \cdots &\mathbf{R}^+_{N,\tau_{\ell-1}}\\
\vdots & \vdots & \vdots & \vdots & \vdots & \vdots & \vdots & \vdots & \vdots &\vdots
\end{array}\right)
\]
from (\ref{samplematrix}) we obtain
\begin{equation}
\label{samplingFormulaGen}
\boldsymbol{\alpha}=\mathbf{R}^+_{\mathbf{b},a}\,\Lc_{\mathrm{samp}}x= 
\sum_{j=1}^N\sum_{n=0}^{\ell-1}\Lc_jx(\tau_n)\mathbf{R}^+_{j,\tau_n}=
\sum_{j=1}^N\sum_{n=0}^{\ell-1}\langle G_{j,\tau_n}\mid \boldsymbol{\alpha}\rangle_{L^2(G)}\mathbf{R}^+_{j,\tau_n}
\end{equation}
In particular we derive that $\rank  \mathbf{R}^+_{\mathbf{b},a}=|G|$ and  that the columns of $\mathbf{R}_{\mathbf{b},a}^+$ form a dual frame of $\{G_{j,\tau_n}\}$, $j=1,2,\ldots, N$, $n=0,1,\ldots,\ell-1$.   Any other dual frame of $\{G_{j,\tau_n}\}$, $j=1,2,\ldots, N$, $n=0,1,\ldots,\ell-1$, in $L^2(G)$ is given by the columns of any left-inverse matrix $\mathbf M$  of the matrix $\mathbf{R}_{\mathbf{b},a}$. All these matrices are expressed as (see \cite{penrose:55}) 
\begin{equation}
\label{todasLI}
    	\mathbf{M}=\mathbf{R}^+_{\mathbf{b},a}+\mathbf{U}[\mathbf{I}_{N\ell}-\mathbf{R}_{\mathbf{b},a}\mathbf{R}^+_{\mathbf{b},a}]
\end{equation}
where $\mathbf{U}$ denotes any arbitrary $|G|\times N\ell$ matrix.

\section{The sampling result}
\label{section3}
Let $\bra{x}= \sum_{s\in G} \alpha_s \,U(s)\bra{a}$  a vector of $\Ac_a$;  applying the isomorphism (\ref{iso}) in (\ref{samplingFormulaGen}) we have
\begin{equation}
\label{samplingResult}
   \bra{x}= \Tc^G_a(\boldsymbol{\alpha})=
   \sum_{j=1}^N\sum_{n=0}^{\ell-1} \Lc_j x(\tau_n)\,\Tc^G_{a}(\mathbf{R}^+_{j,\tau_n})
\end{equation}
The sampling function $\Tc^G_{a}(\mathbf{R}^+_{j,\tau_n})$ in (\ref{samplingResult}) do not have, in principle, any special structure since $\mathbf{R}^+_{\mathbf{b},a}$ does not have it. Similarly we may obtain a sampling formula like (\ref{samplingResult}) for each left-inverse of $\mathbf{R}_{\mathbf{b},a}$ in (\ref{todasLI}).   In the next section we construct specific left-inverses of $\mathbf{R}_{\mathbf{b},a}$ such that their associated sampling  formulas take care of the unitary structure of $\Ac_a$.

\subsection*{$G$-compatible left-inverses} 
We denote by $\mathbf{S}$ the first $|H|$ rows of any left-inverse of the matrix $\mathbf{R}_{\mathbf{b},a}$ taken from (\ref{todasLI}); i.e,
\begin{equation}
\label{SPartialEquation}
\mathbf{S}\,\mathbf{R}_{\mathbf{b},a}=\big(\mathbf{I}_{|H|}\,\, \mathbf{O}_{|H|\times (|G|- |H|)}\big)\,.
\end{equation}
Having in mind the structure of $\mathbf{R}_{\mathbf{b},a}$ we write the $|H|\times N\ell$ matrix $\mathbf{S}$ as
\[
\mathbf{S}=\big(\mathbf{S}_1 \, \mathbf{S}_2 \cdots \mathbf{S}_N\big)
\]
where each block $\mathbf{S}_j$ is a $|H|\times \ell$ matrix denoted by $\mathbf{S}_j=\big(S_j^0\, S_j^1\, \dots\, S_j^{\ell-1}\big)$ where 
$S_j^n\in\CC^{|H|}$ for each  $n=0,1,\dots,\ell-1$ and $j=1,2,\dots,N$. From (\ref{estructuraRCasoGeneral}) and (\ref{SPartialEquation}) we have:
\begin{eqnarray*}
    &&\sum_{j=1}^N \sum_{n=0}^{\ell-1} S_j^n \,R_{b_j,a}(\tau_0^{-1}\tau_n^{-1}H)=\mathbf{I}_{|H|}\\
    &&\sum_{j=1}^N \sum_{n=0}^{\ell-1} S_j^n \,R_{b_j,a}(\tau_k^{-1}\tau_n^{-1}H)=\mathbf{O}_{|H|},\quad k=1,2,\dots,\ell-1\,,
\end{eqnarray*}
or equivalently
\begin{eqnarray}
\label{partialEq2Gen}
    &&\sum_{j=1}^N \sum_{n=0}^{\ell-1} S_j^n \,R_{b_j,a}(\tau_n^{-1}H)=\mathbf{I}_{|H|}\\
\label{partialEq3Gen}
    &&\sum_{j=1}^N \sum_{n=0}^{\ell-1} S_j^n \,R_{b_j,a}(\tau_n^{-1}\tau_k^{-1}H)=\mathbf{O}_{|H|},\quad k=1,2,\dots,\ell-1\,.
\end{eqnarray}

Now, we form the $|G|\times N\ell$ matrix 
$\widetilde{\mathbf{S}}=\big(\widetilde{\mathbf{S}}_1\,\widetilde{\mathbf{S}}_2,\dots\widetilde{\mathbf{S}}_N\big)$. Each
$|G|\times \ell$ block $\widetilde{\mathbf{S}}_j$, $j=1,2,\dots,N$, is formed from the columns of $\mathbf{S}_j$ in the following manner:
\[
   \widetilde{\mathbf{S}}_j:= \left(\begin{array}{cccc}
                          S_j^0 & S_j^1 & \cdots & S^{\ell-1}_j\\
                          S_j^{0,1} & S_j^{1,1} & \cdots & S^{\ell-1,1}_j\\
                          \vdots & \vdots &\cdots &\vdots\\
                          S_j^{0,\ell-1} & S_j^{1,\ell-1} & \cdots & S^{\ell-1,\ell-1}_j
                        \end{array}\right)
\]
where, for $i=1, 2,\dots, \ell-1$ and $n,k=0, 1, \dots ,\ell-1$, we set
\begin{equation}
\label{crucial}
S_j^{n,i}:=S_j^k\quad  \mathrm{whenever} \,\,  \tau_n^{-1}\tau_i^{-1}=\tau_k^{-1} \mathrm{(or, equivalently}, \, \, \tau_i\, \tau_n=\tau_k)\,. 
\end{equation}
\begin{lema}
The above $|G|\times N\ell$ matrix $\widetilde{\mathbf{S}}$ is a left-inverse of $\mathbf{R}_{\mathbf{b},a}$, i.e, $\widetilde{\mathbf{S}}\,\mathbf{R}_{\mathbf{b},a}=\mathbf{I}_{|G|}$.
\end{lema}
\noindent\textit{Proof:}
  From (\ref{partialEq2Gen}) and (\ref{partialEq3Gen}), for each $i=1,2,\dots,\ell-1$, we have
\[
\sum_{j=1}^N \sum_{n=0}^{\ell-1} S_j^{n,i}\,R_{b_j,a}(\tau_n^{-1}\tau_i^{-1}H)=
\sum_{j=1}^N \sum_{k=0}^{\ell-1} S_j^k\,R_{b_j,a}(\tau_k^{-1} H)= \mathbf{I}_{|H|}
\]
and
\[
\sum_{j=1}^N \sum_{n=0}^{\ell-1} S_j^{n,i}\,R_{b_j,a}(\tau_n^{-1}\tau_p^{-1}H)=
\sum_{j=1}^N \sum_{k=0}^{\ell-1} S_j^k\,R_{b_j,a}(\tau_n^{-1}\tau_p^{-1} H)= \mathbf{O}_{|H|},\quad p\neq i\,.
\]
As a consequence, we deduce that $\widetilde{\mathbf{S}}\,\mathbf{R}_{\mathbf{b},a}=\mathbf{I}_{|G|}$.  \hfill$\Box$

We denote the columns of $\widetilde{\mathbf{S}}$ as
\begin{equation}
\label{Stilde}
\widetilde{\mathbf{S}}= \left(\begin{array}{cccccccccc}
\vdots & \cdots &\vdots & \vdots &\cdots &\vdots & \vdots & \vdots&\cdots & \vdots\\
\widetilde{\mathbf{S}}_{1,0} & \cdots &\widetilde{\mathbf{S}}_{1,\ell-1} & \widetilde{\mathbf{S}}_{2,0} & \cdots &\widetilde{\mathbf{S}}_{2,\ell-1} & \cdots & \widetilde{\mathbf{S}}_{N,0} & \cdots&\widetilde{\mathbf{S}}_{N,\ell-1}\\
\vdots & \cdots &\vdots & \vdots &\cdots &\vdots & \vdots & \vdots&\cdots & \vdots
\end{array}\right)
\end{equation}
Using the left-inverse $\widetilde{\mathbf{S}}$ of $\mathbf{R}_{\mathbf{b},a}$ instead of $\mathbf{R}^+_{\mathbf{b},a}$ in (\ref{samplingFormulaGen}), for each 
$\bra{x}= \sum_{s\in G} \alpha_s \,U(s)\bra{a}$  in $\Ac_a$ we obtain
\[
\boldsymbol{\alpha}=\widetilde{\mathbf{S}}\,\Lc_{\mathrm{samp}}=
\sum_{j=1}^N\sum_{n=0}^{\ell-1}\Lc_jx(\tau_n)\,\widetilde{\mathbf{S}}_{j,n}\,.
\]
Therefore,
\[
 \bra{x}= \Tc^G_{a}(\boldsymbol{\alpha})=\sum_{j=1}^N\sum_{n=0}^{\ell-1}\Lc_jx(\tau_n)\,\Tc^G_{a}(\widetilde{\mathbf{S}}_{j,n})\,. 
\]
On the other hand, the columns $\widetilde{\mathbf{S}}_{j,n}$, $j=1,2,\dots,N$ and $n=0,1,\dots,\ell-1$, as vectors of $L^2(G)$ satisfy, by construction, see (\ref{crucial}), the crucial property (for our sampling purposes)
\[
      \widetilde{\mathbf{S}}_{j,n}= L_{\tau_n} \widetilde{\mathbf{S}}_{j,0},\quad j=1,2,\dots,N,\,\, n=0,1,\dots,\ell-1\,.
\]
Hence, the shifting property (\ref{shiftgroup}) gives

\begin{eqnarray*}
\bra{x}&=&\sum_{j=1}^N\sum_{n=0}^{\ell-1}\Lc_jx(\tau_n)\,\Tc^G_{a}(\widetilde{\mathbf{S}}_{j,n})
  =\sum_{j=1}^N\sum_{n=0}^{\ell-1}\Lc_jx(\tau_n)\,\Tc^G_{a}(L_{\tau_n}\widetilde{\mathbf{S}}_{j,0})\\
  &=&\sum_{j=1}^N\sum_{n=0}^{\ell-1}\Lc_jx(\tau_n)\,U(\tau_n) \Tc^G_{a}(\widetilde{\mathbf{S}}_{j,0})\,.
\end{eqnarray*}

Therefore, we have proved that, for any $\bra{x}\in \Ac_a$ the sampling expansion
\begin{equation}\label{sampling_expansion}
   \bra{x}= \sum_{j=1}^N\sum_{n=0}^{\ell-1}\Lc_jx(\tau_n)\,U(\tau_n)\bra{c_j}
\end{equation}
holds, where $\bra{c_j} =\Tc^G_{a}(\widetilde{\mathbf{S}}_{j,0})\in \Ac_a$, $j=1,2,\dots,N$.   Notice that we have obtained a sampling result of the desired form, Eq. (\ref{general_sampling}), where the desired sampling vectors are given by $|C_n,j \rangle = U(\tau_n)|c_j\rangle$ and the generalized samples are obtained by  convolution on $L^2(G)$ with the matrix of cross-covariances.  In fact, collecting all the pieces we have obtained until now we have the following result: 

\begin{teo}
\label{teoGeneral}
Given the $N \ell \times |G|$ matrix  $\mathbf{R}_{\mathbf{b},a}$ defined in (\ref{matrixcc}), the following statements are equivalent:
\begin{enumerate}
\item $\rank \mathbf{R}_{\mathbf{b},a}=|G|$
\item There exists a $|H| \times N\ell$ matrix $\mathbf{S}$ such that 
    \[
      \mathbf{S}\,\mathbf{R}_{\mathbf{b},a}=\big(\mathbf{I}_{|H|}\, \mathbf{O}_{|H|\times (|G|-|H|)}\big)
    \]
\item There exist $N$ vectors $\bra{c_j} \in\Ac_a$, $j=1,2,\dots,N$, such that  
$\{U(\tau_n)\bra{c_j}\}$, $j=1,2,\ldots, N$, $n=0,1,\ldots,\ell-1$ is a frame for  $\Ac_a$, and for any 
$\bra{x}\in \Ac_a$ the expansion 
\[
\bra{x}= \sum_{j=1}^N\sum_{n=0}^{\ell-1}\Lc_jx(\tau_n)\,U(\tau_n)\bra{c_j}
\]
holds.
\item There exists a frame $\{\bra{C_{j,n}}\}$, $j=1, 2,\ldots, N$, $n=0,1,\ldots,\ell-1$, in $\Ac_a$ such that, for each $\bra{x}\in \Ac_a$ the expansion
\[
\bra{x}= \sum_{j=1}^N\sum_{n=0}^{\ell-1}\Lc_jx(\tau_n)\, \bra{C_{j,n}}
\]
holds.
\end{enumerate}
\end{teo}
\noindent\textit{Proof:}
That condition $(1)$ implies condition $(2)$ and condition $(2)$ implies condition $(3)$ have been proved above. Obviously, condition $(3)$ implies condition $(4)$: take $\bra{C_{j,n}}=U(\tau_n)\bra{c_j}$ for $j=1, 2, \dots, N$ and $n=0, 1, \dots, \ell-1$. Finally, as a consequence of Corollary \ref{cor1}, condition $(4)$ implies condition $(1)$.  \hfill$\Box$

For the particular case where $N=|H|$ we obtain:
\begin{cor}
Assume that $N=|H|$ and consider the $|G|\times |G|$ matrix of cross-covariances $\mathbf{R}_{\mathbf{b},a}$ defined in (\ref{matrixcc}). The following statements are equivalent:
\begin{enumerate}
\item The  matrix $\mathbf{R}_{\mathbf{b},a}$ is invertible.
\item There exist $|H|$ unique elements $\bra{c_j}\in \Ac_a$, $j=1, 2, \dots, |H|$, such that the sequence $\big\{U(\tau_n) \bra{c_j} \big\}$, $j=1,2,\ldots, |H|$, $n=0,1,\ldots,\ell-1$ is a basis for $\Ac_a$, and  the expansion of any $\bra{x}\in  \Ac_a$ with respect to this basis is
\[
\bra{x}= \sum_{j=1}^{|H|} \sum_{n=0}^{\ell-1} \Lc_j x(\tau_n)\, U(\tau_n) \bra{c_j} \,.
\]
\end{enumerate}

In case the equivalent conditions are satisfied, the interpolation property $\Lc_{j}c_{j'}(\tau_n)=\delta_{j,j'}\,\delta_{n,0}$, whenever $n=0,1,\dots ,\ell-1$ and  $j,j'=1,2, \dots, |H|$, holds.
\end{cor}
\noindent\textit{Proof:}
Notice that the inverse matrix $\mathbf{R}_{\mathbf{b},a}^{-1}$ has necessarily the structure of the matrix $\widetilde{\mathbf{S}}$ in (\ref{Stilde}). The uniqueness of the expansion with respect to a basis gives the interpolation property.   \hfill$\Box$

\subsection*{Sampling the dynamics}

So far, the group $G$ has played no dynamical role, that is, if the dynamical evolution of the state $\bra x$ is given by the one-parameter group of unitary operatos $U_t = \exp (-it\sf{H})$ defined by the Hamiltonian operator $\sf{H}$, no assumption on the commutation relations of the operators $U(g)$ and $\sf{H}$ is made.  However, if we assume that $G$ is a symmetry group of the system, that is $[U(g), \sf{H}] = 0$ for all $g\in G$, then the evolution $x(t)$ of the initial state takes a particularly simple form.   Actually $\bra{x(t)} = U_t \bra{x}$, then using Eq. (\ref{sampling_expansion}), we get:
\[
\bra{x(t)} = U_t \bra{x}= \sum_{j=1}^N\sum_{n=0}^{\ell-1}\Lc_jx(\tau_n)\,U(\tau_n) U_t\bra{c_j} = \sum_{j=1}^N\sum_{n=0}^{\ell-1}\Lc_jx(\tau_n)\,U(\tau_n)\bra{c_j(t)} \, ,
\]
where $\bra{c_j(t)} $ denotes the dynamical evolution of the states $\bra{c_j}$, $j=1, 2,\ldots, N$.  Thus the evolved state $\bra{x(t)}$ can be recovered from the initial samples and the evolved states $\bra{c_j(t)}$, $j=1, 2,\ldots, N$.

\section{Some simple examples}
\label{section4}

We will end up the discussion by illustrating the obtained results with two simple examples: the cyclic group $\mathbb{Z}_N$ and the dihedral group $D_3$, or the $C_{3v}$ group in the molecular symmetry notation, which is the symmetry group of molecules such as ammonia or phosphorus oxycloride.

\subsection*{The cyclic group case}

This case corresponds to that of a unitary operator $U:\Hc \rightarrow \Hc$ in a Hilbert space $\Hc$ such that for some $\bra{a}\in  \Hc$ there exists $M\in \NN$  satisfying $U^M\bra{a}=\bra{a}$ and the set $\big\{\bra{a}, U\bra{a}, U^2\bra{a},\dots,U^{M-1}\bra{a}\big\}$ is linearly independent in $\Hc$. 

If we consider a positive integer $r$ such that $r$ divides $M$ and denote $\ell=M/r$, the goal is to obtain finite frames in the subspace $\Ac_a=\big\{\sum_{k=0}^{M-1} \alpha_k U^k a\,:\, \alpha_k\in \CC  \big\}$ of $\Hc$, having the form $\big\{U^{rn} \bra{c_j}\big\}$, $j=1,2,\ldots, N$, $n=0,1,\ldots,\ell-1$ where $\bra{c_j}\in \Ac_a$, $j=1,2,\dots, N$, such that  any $\bra{x}\in   \Ac_a$ can be recovered from the samples $\big\{\Lc_jx(rn):=\langle U^{rn}b_j \mid x \rangle_\Hc\big\}$, $j=1,2,\ldots, N$, $n=0,1,\ldots,\ell-1$, by means of the sampling expansion
\[
\bra{x}= \sum_{j=1}^N \sum_{n=0}^{\ell-1} \Lc_j x(rn)\, U^{rn} \bra{c_j}\,,
\]
which takes care of the $U$-structure of $\Ac_a$. In this example we are considering the cyclic group $G:=\ZZ_M$, the unitary representation $n\mapsto U(n):=U^n$ and its cyclid subgroup $H:=\ZZ_r$. This particular case has been deeply studied in Refs. \cite{hector:15,garcia:15}.

\medskip

A practical example taken from {\em Signal Processing} consists of the periodic extension of finite signals. Indeed, consider the space $\ell^2_M(\ZZ)$ of $M$-periodic signals with inner product $\langle \mathbf{x}\mid \mathbf{y} \rangle_{\ell^2_M}=\sum_{m=0}^{M-1} \overline{x(m)}\, y(m)$. If we take, for instance, the $M$ periodic signal $\mathbf{a}:=(1, 0,\dots , 0)$  and the cyclid shift operator 
\[
U: \mathbf{x}=\big\{x(m)\big\} \longmapsto U\mathbf{x}:=\big\{x(m-1)\big\}
\] 
in $\ell^2_M(\ZZ)$, we trivially obtain that $U^M \mathbf{a}=\mathbf{a}$ and $\Ac_{\mathbf{a}}=\ell^2_M(\ZZ)$.

Fixed $N$ signals $\mathbf{b}_j\in \ell^2_M$, $j=1,2,\dots, N$, each sample of $\mathbf{x} \in \ell^2_M$ is obtained from the $M$-periodic convolution
\[
\Lc_j \mathbf{x}(rn):=\langle U^{rn}\mathbf{b}_j\mid \mathbf{x} \rangle_{ \ell^2_N}=\sum_{m=0}^{M-1}x(m)\,\overline{\mathbf{b}_j(m-rn)}=
\big(\mathbf{x}*\mathbf{h}_j\big)(rn)\,, \quad n=0, 1, \dots, \ell\,,
\]
where $\mathbf{h}_j(m)=\overline{\mathbf{b}_j(-m)}$, \,\,$m=0,1, \dots, M-1$.

As the cross-covariance $R_{b_j,a}(m)=\langle  \mathbf{b}_j \mid U^m \mathbf{a}\rangle_{\ell^2_N}=\overline{b_j(m)}$, each 
$ \ell \times M$ block $\mathbf{R}_{b_j,a}$, $j=1,2,\dots, N$, of the $N\ell \times M$ matrix $\mathbf{R}_{\mathbf{b},a}$ in (\ref{matrixcc}) takes the form
\[
\mathbf{R}_{b_j,a}= \left(\begin{array}{cccc}
\overline{b_j(0)} & \overline{b_j(1)} & \cdots & \overline{b_j(M-1)} \\
      \overline{b_j(M-r)} & \overline{b_j(M-r+1)}& \cdots & \overline{b_j(2M-r-1)} \\
           \vdots & \vdots& \ddots & \vdots \\
                \overline{b_j(M-r(\ell-1))} &  \overline{b_j(M-r(\ell-1)+1)}& \cdots &  \overline{b_j(2M-1-r(\ell-1))}\\
\end{array}\right)
\]
where the sampling period $r$ divides $M$ and $\ell=M/r$. In case the rank of $\mathbf{R}_{\mathbf{b},a}$ is $M$, which implies $N\geq r$ convolution systems, from Theorem \ref{teoGeneral} we obtain in $\ell^2_M(\ZZ)$ a sampling formula as
\[
\mathbf{x}(m)=\sum_{j=1}^N \sum_{n=0}^{\ell-1} \Lc_j \mathbf{x}(rn)\,\mathbf{c}_j(m-rn) 
=\sum_{j=1}^N \sum_{n=0}^{\ell-1} \big(\mathbf{x}*\mathbf{h}_j\big)(rn)\,\mathbf{c}_j(m-rn)\,,
\]
where $m=0, 1, \dots, M-1$. The sampling sequences in $\ell^2_M(\ZZ)$ are $\mathbf{c}_j=\mathcal{T}_{a}^G\big(\widetilde{\mathbf{S}}_{j,0}\big)$, $j=1,2,\dots, s$, where $\widetilde{\mathbf{S}}_{j,0}$ are the corresponding columns of any structured left-inverse $\widetilde{\mathbf{S}}$ of $\mathbf{R}_{\mathbf{b},a}$  as in (\ref{Stilde}).
Note that $\mathbf{c}_j$ is nothing but the $M$-periodic sequence in $\ell^2_M(\ZZ)$ derived from the column
$\widetilde{\mathbf{S}}_{j,0}\in \CC^M$ of $\widetilde{\mathbf{S}}$.
\subsection*{The dihedral group case}

Let $D_3$ be the dihedral group $D_3=\big\{e,\, g,\, g^2,\, \tau,\, \tau g,\, \tau g^2\big\}$ of the symmetries of the equilateral triangle, i.e., $g$ is a rotation of $2\pi/3$ and $\tau$ is the axial reflection. They satisfies the relationships $g^3=\tau^2=e$ and $\tau g=g^2\tau$. Consider its normal subgroup $H=\big\{e,\, g,\,g^2\big\}$; as a consequence, the quotient group $D_3/H=\{[e],[\tau]\}$, is a cyclic group of order $2$, and $K=\{e, \tau\}$.

Let $s\in D_3 \mapsto U(s)\in \Uc(\Hc)$ be a unitary representation of $D_3$ in a Hilbert space $\Hc$. Fixed $\bra{a}\in  \Hc$ we consider $\Ac_a$ the  subspace of $\Hc$ spanned by $\big\{U(s)a\,:\, s\in D_3\big\}$. In case this set is linearly independent in $\Hc$ it can be described as the unique expansion $\bra{x}= \sum_{s\in D_3} \alpha_s   \,U(s)\bra{a} $, with $\alpha_s  \in \CC$.

Assume that $N$ $\Lc_j$-systems, $j=1, 2, \dots, N$, are defined on $\Ac_a$. In this case, each $2\times 6$ block $\mathbf{R}_{b_j,a}$ of the $2N\times 6$ matrix $\mathbf{R}_{\mathbf{b},a}$ in (\ref{matrixcc}) is given by
\[
  \mathbf{R}_{b_j,a}=
      \left(\begin{array}{cccccc}
      R_{b_j,a}(e)&R_{b_j,a}(g) &R_{b_j,a}(g^2)& R_{b_j,a}(\tau)& 
      											R_{b_j,a}(\tau g)& R_{b_j,a}(\tau g^2)\\
      R_{b_j,a}(\tau)&R_{b_j,a}(\tau g) &R_{b_j,a}(\tau g^2)& R_{b_j,a}(e)& 
      											           R_{b_j,a}(g)& R_{b_j,a}(g^2)
      \end{array}\right)
\]
If $\rank \mathbf{R}_{\mathbf{b},a}=6$, then, according to Theorem \ref{teoGeneral} there exist $N$ vectors $\bra{c_j}\in \Ac_a$, $j=1, 2, \dots, N$, such that the sequence $\big\{U(s) \bra{c_j}\big\}$, $j=1,2,\ldots, N$, $s\in \{e, \tau \}$ is a frame for $\Ac_a$ and, for any $\bra{x}\in     \Ac_a$ the sampling expansion
\begin{equation*}
 \bra{x}= \sum_{j=1}^N \sum_{s\in\{e,\tau\}}\Lc_jx(s)\,U(s)\bra{c_j}\,,
\end{equation*}
holds. Moreover, $\bra{c_j}=\Tc_{a}^{D_3}(\widetilde{\mathbf{S}}_{j,e})$, $j=1,2,\dots, N$, where $\widetilde{\mathbf{S}}_{j,e}$ denotes the corresponding column of any $6\times 2N$ left-inverse $\widetilde{\mathbf{S}}$ of $\mathbf{R}_{\mathbf{b},a}$ as in (\ref{Stilde}).   A similar result applies for the general dihedral group $D_m$.
\section{Discussion and conclusions}
\label{section5}

A sampling expansion for a vector state $\bra{x}$ in an invariant subspace  $\Ac_a$ of a Hilbert space $\mathcal{H}$ with respect a unitary representation $U(g)$ of a given finite group $G$ has been obtained.    The generalized samples are obtained by convolution in the auxiliary Hilbert space $L^2(G)$, of the the fundamental cross-covariance matrix $\mathbf{R}_{\mathbf{b},a}$ of the finite sequence of vectors $U(\tau_n) \bra{b_j}$ with the original state.  The reconstruction is achieved as a linear superposition of a finite frame $\big\{\bra{C_{j,n}}\big\}$ with these coefficients.  The elements $\bra{C_{j,n}}$ of the frame  are compatible with the unitary structure of the problem in the sense that they have the form $\bra{C_{j,n}}=U(\tau_n) \bra{c_j}$ for some elements $\bra{c_j}$ in $\Ac_a$ and they are constructed in a natural way from a left-inverse of the fundamental cross-covariance matrix.   

The sampling expansions obtained would allow for a reconstruction of the original state, as well as its unitary evolution if the group $G$ is a symmetry of the dynamics, using frames adapted to the structure of the problem, that is, a family of states which, in general, do not form an orthonormal basis of the system and prepared independently of the initial state by using the geometry of the group $G$.    The determination of the samples by means of generalized measurements on the auxiliary Hilbert space $L^2(G)$ and the reconstruction of the unitary evolution will be discussed in subsequent works. 

The use of frames adapted to the geometry of the group $G$ and an Abelian subgroup $H$ as discussed in this paper, could be relevant in  solving such relevant problem as Kitaev's Abelian stabilizer problem or designing faster phase estimation quantum algorithms \cite{Ki97}. 

Molecular symmetry theory will constitute another obvious applications of the theory as it would provide a new way of representing molecular states by using frames and generalized samples. Such issues will be considered in future publications.

The results obtained in this paper for finite groups have natural extensions to compact and type I discrete groups.  Other groups of physical relevance like the Heisenberg-Weyl group and other nilpotent and solvable groups will be discussed elsewhere.  

\medskip

\noindent{\bf Acknowledgments:} 
This work has been supported by the grant MTM2014-54692-P from the Spanish {\em Ministerio de Econom\'{\i}a y Competitividad (MINECO)} and and QUITEMAD+, S2013/ICE-2801.

\vspace*{0.3cm}

\end{document}